\documentclass[cits]{PoS}

\newcommand{\myangle}{-90}
\newcommand{\mywidth}{162pt}

\title{Hadronic contribution to g-2 from twisted mass fermions}

\ShortTitle{Hadronic contribution to g-2 from twisted mass fermions}

\author{\speaker{Dru Renner}\footnote{On behalf of the European Twisted Mass Collaboration.}\,\,$^a$ and Xu Feng$^{a, b}$\\
\llap{$^a$}DESY, Platanenallee 6, D-15738 Zeuthen, Germany\\
\llap{$^b$}Universit\"at M\"unster, Institut f\"ur Theoretische Physik, Wilhelm-Klemm-Strasse 9, D-48149 M\"unster, Germany\\
E-mail: \email{dru.renner@desy.de}, \email{xu.feng@desy.de}}

\abstract{We calculate the vacuum polarization tensor for pion masses
from 480 MeV to 270 MeV using dynamical twisted mass fermions at a
lattice spacing of 0.086 fm.  We analyze the form of the
polarization tensor on the lattice using the symmetries of twisted
QCD and we study both finite size effects and lattice artifacts at a
pion mass of $310~\mathrm{MeV}$.  Results for the lowest order
hadronic contribution to g-2 are presented and the impact of
systematic errors is discussed.
\vspace{65pt}
\begin{center}
\includegraphics[width=100pt]{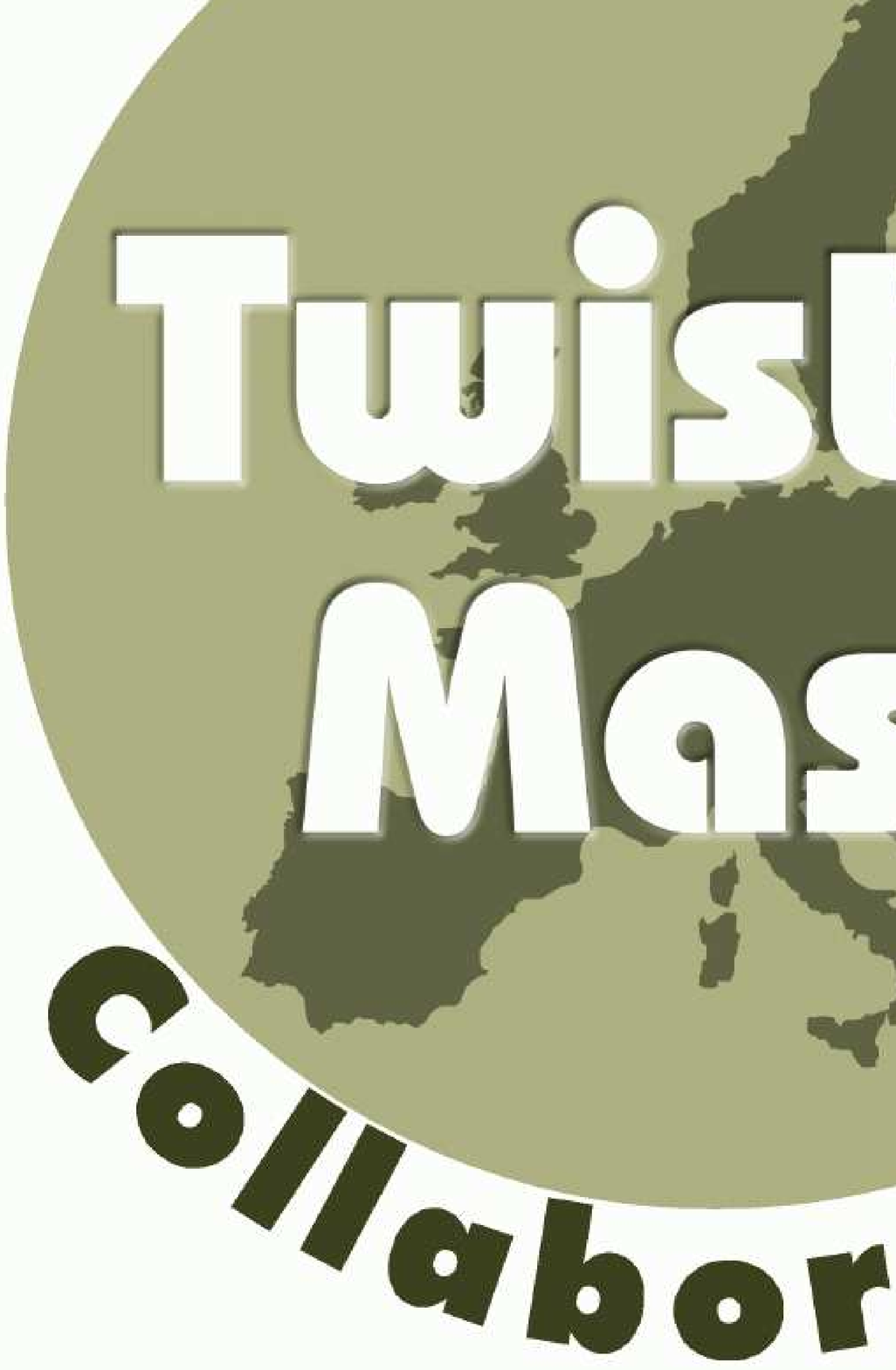}
\end{center}
}

\FullConference{The XXVI International Symposium on Lattice Field Theory\\
July 14 - 19, 2008\\
Williamsburg, Virginia, USA}

\begin{document}

\section{Introduction}

The anomalous magnetic moment of the muon, $a_\mu = (g_\mu-2)/2$, has
been both measured and calculated to high precision, possibly
revealing a small discrepancy between Nature and the Standard Model.
The measurement of the Muon (g-2) Collaboration is
$a_\mu^{\mathrm{ex}}=11\,659\,208.0(6.3) ×
10^{-10}$~\cite{Bennett:2006fi} and has a fractional accuracy
of $0.54\cdot 10^{-6}$.  The Standard Model value has been estimated
by many authors.  One recent review~\cite{Jegerlehner:2007xe} quotes a
value of $a_\mu^{\mathrm{th}}=11\,659\,179.3(6.8) × 10^{-10}$, which
has just a slightly higher fractional error of $0.58\cdot 10^{-6}$.
This results in a discrepancy of $3.1\,\sigma$.  Other theoretical
estimates produce a range of discrepancies from $0.9\,\sigma$ to
$3.4\,\sigma$\footnote{The range of values is determined by examining
the references given in~\cite{Jegerlehner:2007xe}.  The dominant
source of this variation is the discrepancy between
$e^{+}e^{-}$ and $\tau$ data used to determine
$a_\mu^{\mathrm{had}}$.}, but in all cases the dominant source of
error for the Standard Model calculation is the leading order (in the
QED coupling) hadronic contribution, $a_\mu^{\mathrm{had}}$.  The
value quoted in~\cite{Jegerlehner:2007xe},
$a_\mu^{\mathrm{had}}=692.1(5.6) × 10^{-10}$,
alone represents $60\%$ of the theoretical error.  This quantity is a
pure QCD observable and has been shown to be calculable in lattice QCD
calculations even in Euclidean space~\cite{Blum:2002ii}.

This hadronic contribution to $a_\mu$ is the focus of this work.  We
present our initial calculation of $a_\mu^{\mathrm{had}}$ using two-flavor 
maximally twisted mass fermions.  This is only the second
full QCD calculation of this quantity and represents the first such
calculation to examine finite size effects and lattice
artifacts.  As we demonstrate, cleanly controlling all sources of
systematic error will be very important to reliably calculate
$a_\mu^{\mathrm{had}}$.

\section{Calculation}

The leading order hadronic contribution due to vacuum polarization is
\begin{equation}
a_\mu^{\mathrm{had}} = \alpha^2 \int_{0}^{\infty}\frac{dq^2}{q^2} w(q^2/m_\mu^2)( \pi(q^2)-\pi(0))
\label{eq_amu}
\end{equation}
where $m_\mu$ is the muon mass and $w(q^2/m_\mu^2)$ is given
in~\cite{Blum:2002ii}.  The vacuum polarization, $\pi(q^2)$, is
determined from the vacuum polarization tensor, $\pi_{\mu\nu}(q)$, by
\begin{equation}
\pi_{\mu\nu}(q) = \int\!\! d^4\!x\, e^{iq\cdot(x-y)} \langle J_\mu(x) J_\nu(y) \rangle = (q_\mu q_\nu-q^2\delta_{\mu\nu})\pi(q^2)
\label{eq_pimunu}
\end{equation}
where $J_\mu(x)$ is the electromagnetic quark current.  In particular,
we note that the momentum integral in Eq.~\ref{eq_amu} is performed for $q^2>0$~\cite{Blum:2002ii}, thus $\pi(q^2)$ can be calculated directly
from lattice QCD.  Furthermore, we remark that the momentum integral
is peaked at small momentum and the kernel, $w(q^2/m_\mu^2)$, attains
a maximal value at $q^2=(\sqrt{5}-2)m_\mu^2 \approx
0.003~\mathrm{GeV}^2$.  (The inverse power of $q^2$ is canceled by the
subtraction $\pi(q^2)-\pi(0)$, which is proportional to $q^2$.)
Meanwhile the smallest momentum accessible in our finite volume
calculation is $q^2=(2\pi/L)^2 \approx 0.05~\mathrm{GeV}^2$.
Therefore a reliable low $q^2$ extrapolation is essential to
calculate $\pi(q^2)$.  In particular the ultra-violet subtraction at
$q^2=0$ required to renormalize $\pi(q^2)$ induces larger
uncertainties than naively expected.

\section{Lattice Details}

We calculate $\pi_{\mu\nu}(q^2)$ using dynamical maximally twisted
mass fermions.  The twisted quark mass provides an infra-red regulator
that bounds the determinant of the fermion action hence eliminating
exceptional configurations~\cite{Frezzotti:1999vv,Frezzotti:2000nk}.
Additionally, at maximal twist physical observables are automatically
accurate to ${\cal O}(a^2)$ in the lattice
spacing~\cite{Frezzotti:2003ni}.

The flavor diagonal currents retain their usual form undering twisting.\footnote{This
follows simply from $Q \gamma_\mu = \exp(-i\gamma_5\tau_3\theta) Q
\gamma_\mu \exp(-i\gamma_5\tau_3\theta)$ for $Q=1$ and $\tau_3$.}
Additionally, in the twisted
basis we can use the conserved Noether current instead of the local
current.  This eliminates the renormalization factor required for the
local current and ensures that the Ward identity holds even
for non-zero lattice spacing.  The conserved current in the twisted
basis is given as follows,
\begin{displaymath}
J^\mathrm{tw}_{\mu x} = \frac{1}{2}\left\{ \bar{\chi}^\mathrm{tw}_{x+\hat{\mu}}(r+\gamma_\mu)U^\dag_{\mu,x}\chi^\mathrm{tw}_{x} - \bar{\chi}^\mathrm{tw}_{x}(r-\gamma_\mu)U_{\mu,x}\chi^\mathrm{tw}_{x+\hat{\mu}} \right\}
\end{displaymath}
and has the same point-split form as the standard Wilson current.
This can be understood easily once one realizes that both the Wilson mass
term and the twisted mass term are invariant under local QED gauge transformations and hence
do not contribute to the Noether construction of the conserved
current.

The calculation of $\pi_{\mu\nu}(q^2)$ proceeds as for Wilson and
domain wall fermions.  Propagators from point sources at a single site
and the four forward neighbors are calculated and used to construct
the current-current correlator in Eq.~\ref{eq_pimunu}.  The one
exception is that separate $u$ and $d$ quark inversions must be
performed due to the modified $\gamma_5$-hermiticity:\ $\gamma_5
D_u^\dagger \gamma_5 = D_d$.\footnote{This can be seen from the basic
loop expression $\gamma_\mu D^{-1}_u(x,y) \gamma_\nu
D_u^{-1\dagger}(y,x)
= \gamma_\mu D^{-1}_u(x,y) \gamma_\nu \gamma_5
D_d^{-1}(x,y) \gamma_5$.}

\begin{table}
\begin{center}
\begin{tabular}{|c|c|c|c|c|c|c|c|} \hline
$\beta$ & $a\mu$ & $V/a^4$ & $a$           & $L$           & $m_\pi$        & $m_\pi L$ & $N_\mathrm{traj}$ \\ \hline
3.9  & 0.0100 & $24^3\times 48$ & 0.086 & 2.1 & 480 & 5.0 & 120 \\ \hline
3.9  & 0.0085 & $24^3\times 48$ & 0.086 & 2.1 & 450 & 4.7 & 207 \\ \hline
3.9  & 0.0064 & $24^3\times 48$ & 0.086 & 2.1 & 390 & 4.1 & 139 \\ \hline
3.9  & 0.0040 & $24^3\times 48$ & 0.086 & 2.1 & 310 & 3.3 & 178 \\ \hline
3.9  & 0.0030 & $32^3\times 64$ & 0.086 & 2.7 & 270 & 3.7 & 101 \\ \hline
3.9  & 0.0040 & $32^3\times 64$ & 0.086 & 2.7 & 310 & 4.3 & 124 \\ \hline
4.05 & 0.0030 & $32^3\times 64$ & 0.067 & 2.1 & 310 & 3.3 & 104 \\ \hline
\end{tabular}
\end{center}
\caption{Parameters used in this work.  The values of $a$ and $L$ are given in $\mathrm{fm}$ and $m_\pi$ is given in $\mathrm{MeV}$.}
\label{en}
\end{table}

Twisted fermions break flavor symmetry.  However, the $\gamma_5$-hermiticity relates $u$ and $d$ quark loops and
results in $\pi^d_{\mu\nu}(x,y) = \pi^{u\ast}_{\mu\nu}(x,y)$.  This
expression is true for each gauge field configuration.  The
consequence is that $\mathrm{re}(\pi^d(q^2)) = \mathrm{re}(\pi^u(q^2))$.
Hence by simply taking the real part of $\pi(q^2)$, which is real in
the continuum limit, we eliminate any explicit flavor breaking in the
valence sector.
Additionally, we expect
the real part of $\pi$ to be accurate to ${\cal O}(a^2)$, even if the
imaginary part has ${\cal O}(a)$ corrections.

In this work we use the two-flavor dynamical gauge field configurations
from the European Twisted Mass Collaboration~\cite{Boucaud:2007uk,Boucaud:2008xu,Urbach:2007rt}.  The details of the
ensembles used are given in Tab.~\ref{en}.  Additionally, the hadronic
contribution to $a_\mu$ has previously been calculated using quenched
domain wall fermions~\cite{Blum:2002ii}, quenched improved Wilson
fermions~\cite{Gockeler:2003cw} and dynamical rooted asqtad improved
staggered fermions~\cite{Aubin:2006xv}.

\section{Results}

\begin{figure}
\begin{minipage}{210pt}
\includegraphics[width=\mywidth,angle=\myangle]{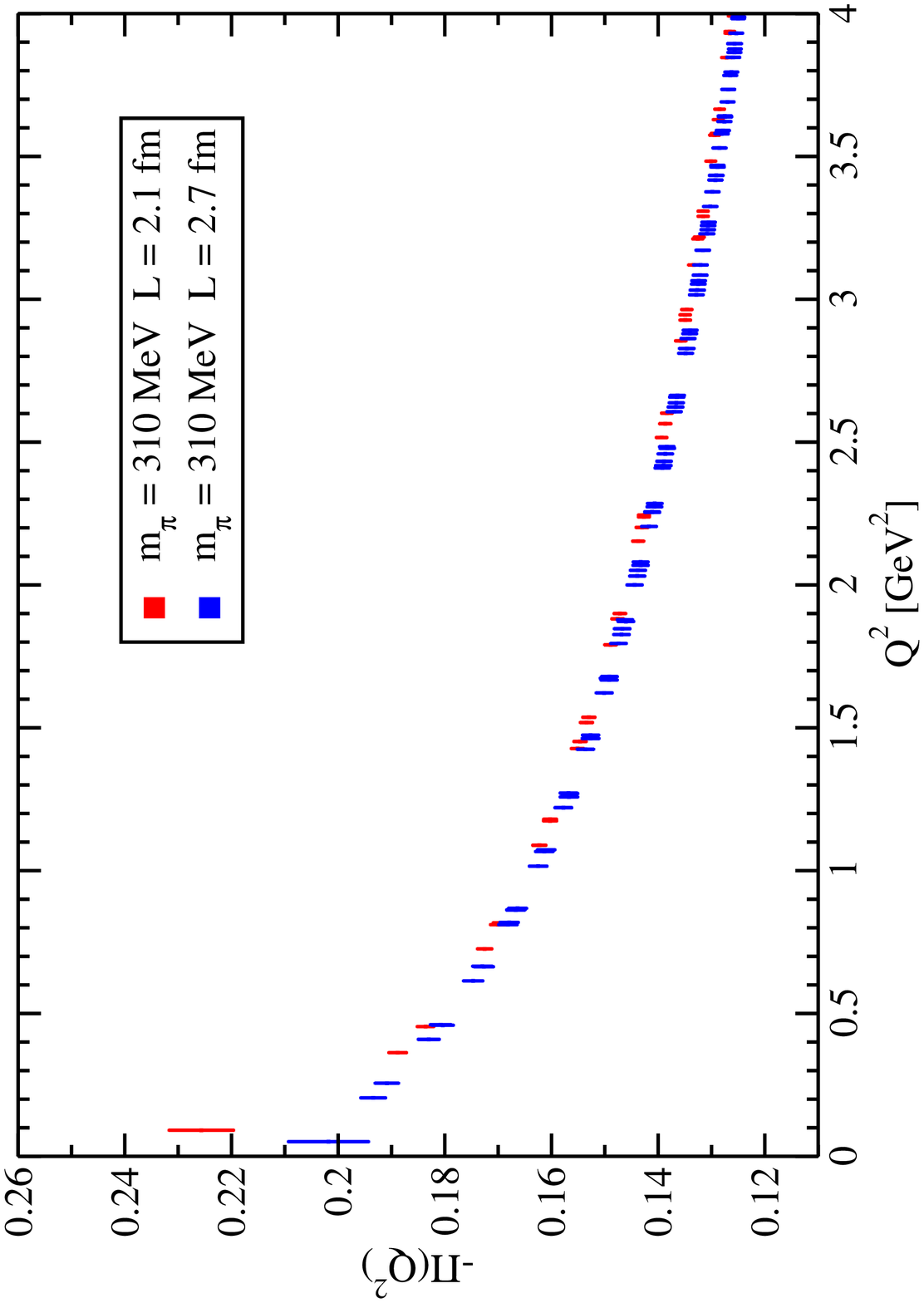}
\caption{Volume dependence of $-\pi(q^2)$.  This quantity requires an ultra-violet subtraction but is infra-red finite.  With the exception of the
lowest $q^2$ point, there is no noticeable finite size effect.}
\label{v_dep_1}
\end{minipage}
\hspace{4pt}
\begin{minipage}{210pt}
\includegraphics[width=\mywidth,angle=\myangle]{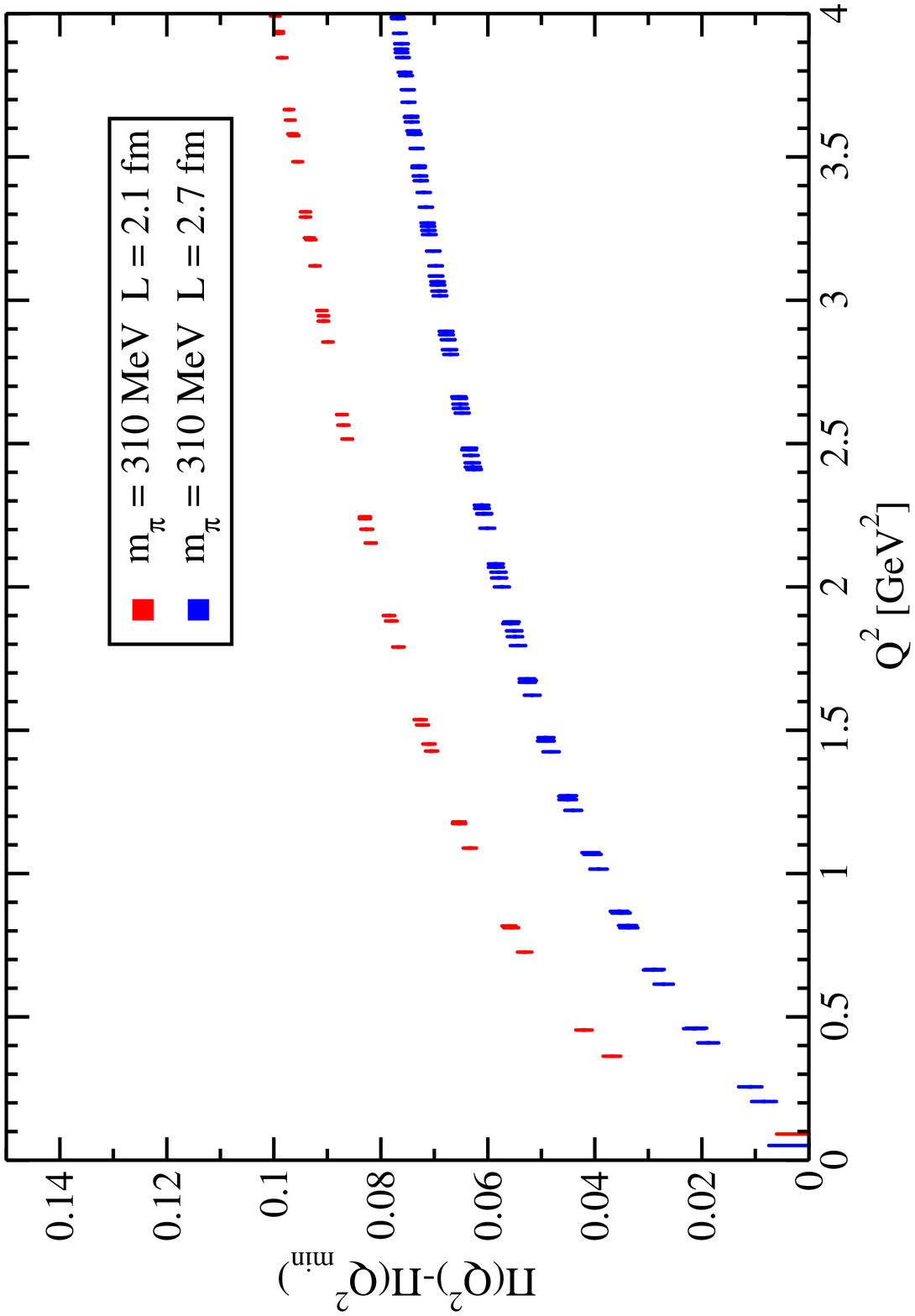}
\caption{Volume dependence of $\pi(q^2)-\pi(0)$.  As an illustration, this quantity is renormalized at the lowest momentum accessible in each volume,
demonstrating the sensitive nature of the subtraction.}
\label{v_dep_2}
\end{minipage}
\end{figure}

First we examine the finite size effects in $\pi(q^2)$.  We have
calculated it at two volumes, $L=2.1~\mathrm{fm}$ and
$L=2.7~\mathrm{fm}$, both with $m_\pi=310~\mathrm{MeV}$ and
$a=0.086~\mathrm{fm}$.  In Fig.~\ref{v_dep_1} we show the results for
$\pi(q^2)$ for these two volumes.  Although $\pi(q^2)$ is an ultra-violet divergent
quantity, for a fixed $a$ we expect a finite large volume
limit.  Fig.~\ref{v_dep_1} demonstrates this rather clearly for all
but the lowest $q^2$ point.  However, the ultra-violet subtraction
required to form $a_\mu$ exaggerates any differences at low $q^2$.
Fig.~\ref{v_dep_2} illustrates this point.  Determining $\pi(0)$
requires a fit, so for the purposes of illustration we perform the
subtraction at the lowest $q^2$ value available for each of the two
volumes.  Fig.~\ref{v_dep_2} demonstrates that this subtraction can
have a large effect on the resulting renormalized quantity.  However,
the integral in Eq.~\ref{eq_amu} is dominated by the region near
$q^2\approx 0.003~\mathrm{GeV}^2$ and hence is not fully sensitive to
the overall shift in Fig.~\ref{v_dep_2}.

\begin{figure}
\begin{minipage}{210pt}
\includegraphics[width=\mywidth,angle=\myangle]{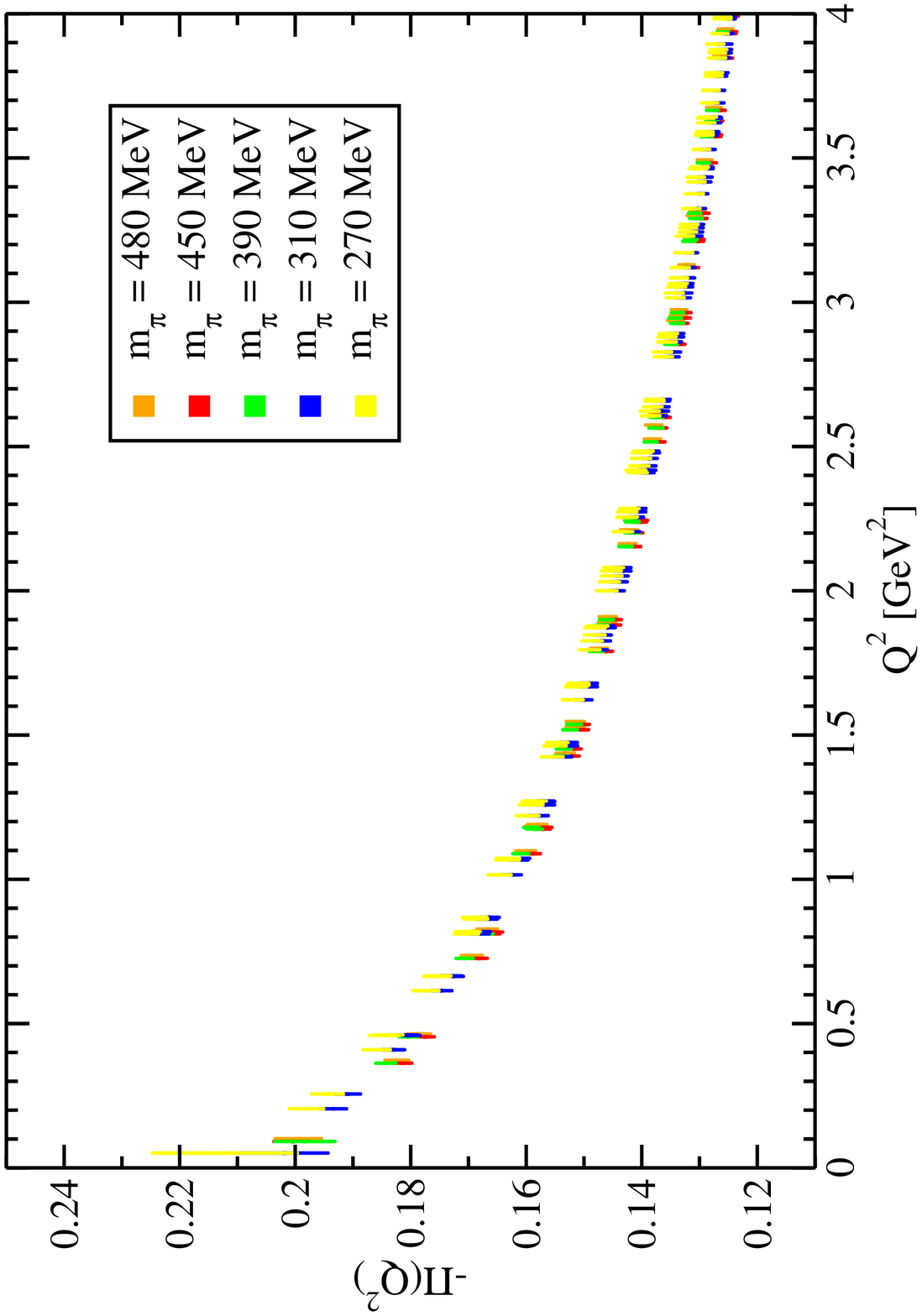}
\caption{Quark mass dependence at large $q^2$.  There is no noticeable quark mass dependence at large $q^2$, consistent
with perturbative QCD expectations.}
\label{mpi_dep_1}
\end{minipage}
\hspace{4pt}
\begin{minipage}{210pt}
\includegraphics[width=\mywidth,angle=\myangle]{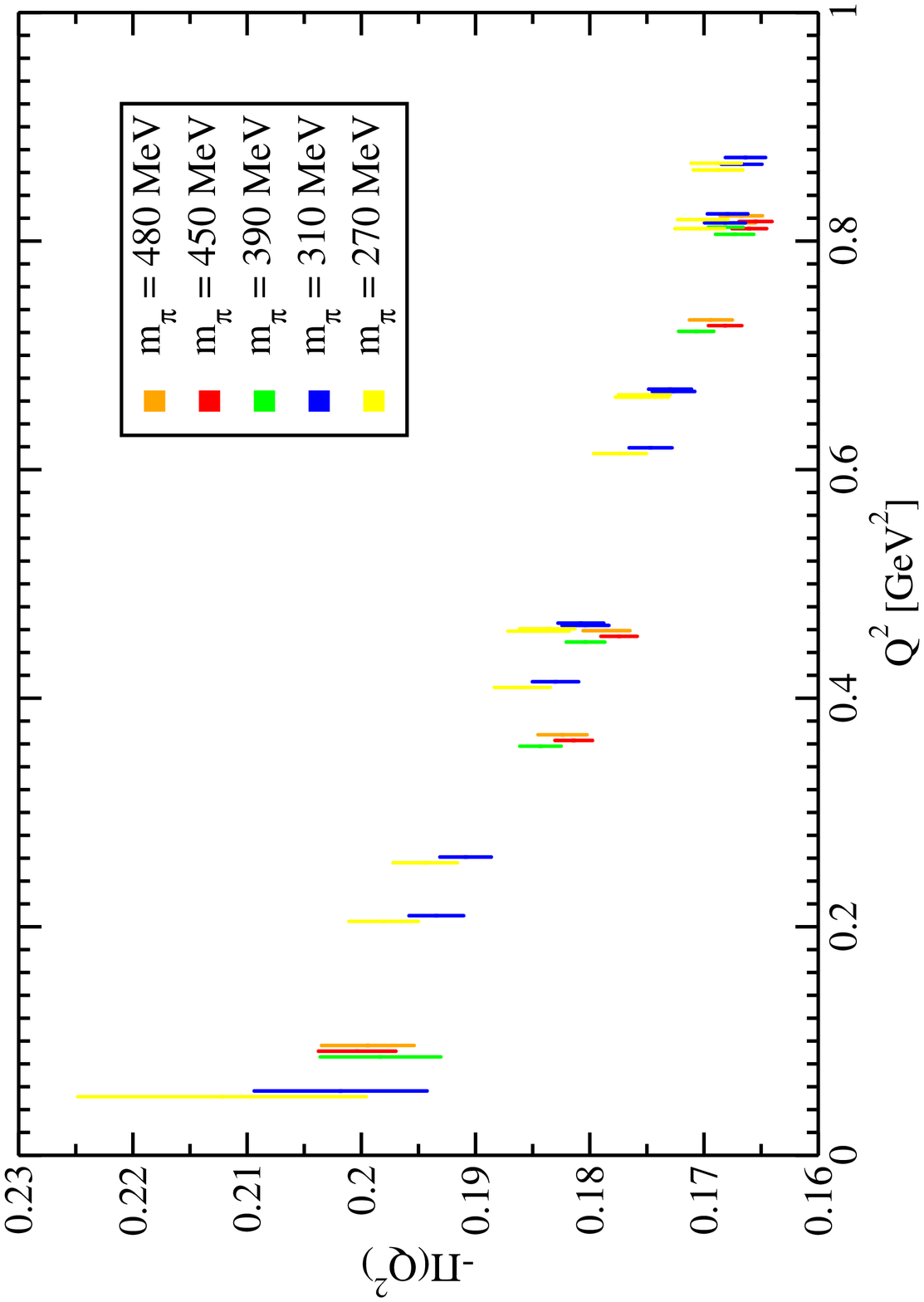}
\caption{Quark mass dependence at low $q^2$.  There is a systematic, but not statistically significant, shift with quark mass
from $m_\pi=450~\mathrm{MeV}$ to $270~\mathrm{MeV}$.}
\label{mpi_dep_2}
\end{minipage}
\end{figure}

Next we study the $m_\pi$ dependence.  Figs.~\ref{mpi_dep_1} and
\ref{mpi_dep_2} show all five values of $m_\pi$ for
$a=0.086~\mathrm{fm}$.  In the case of $m_\pi=310~\mathrm{MeV}$, only
the larger $L=2.7~\mathrm{fm}$ results are shown.
Fig.~\ref{mpi_dep_1} demonstrates that there is no visible quark mass
dependence for large $q^2$, as expected from perturbation theory.  Any
quark mass dependence should be more visible for the low $q^2$ region shown
in Fig.~\ref{mpi_dep_2}.  The error bars are too large to identity any
quark mass dependence, however, the results do appear to systematically
increase when proceeding from the $m_\pi=450~\mathrm{MeV}$ calculation
down to the $270~\mathrm{MeV}$ result.

\begin{figure}
\begin{minipage}{210pt}
\includegraphics[width=\mywidth,angle=\myangle]{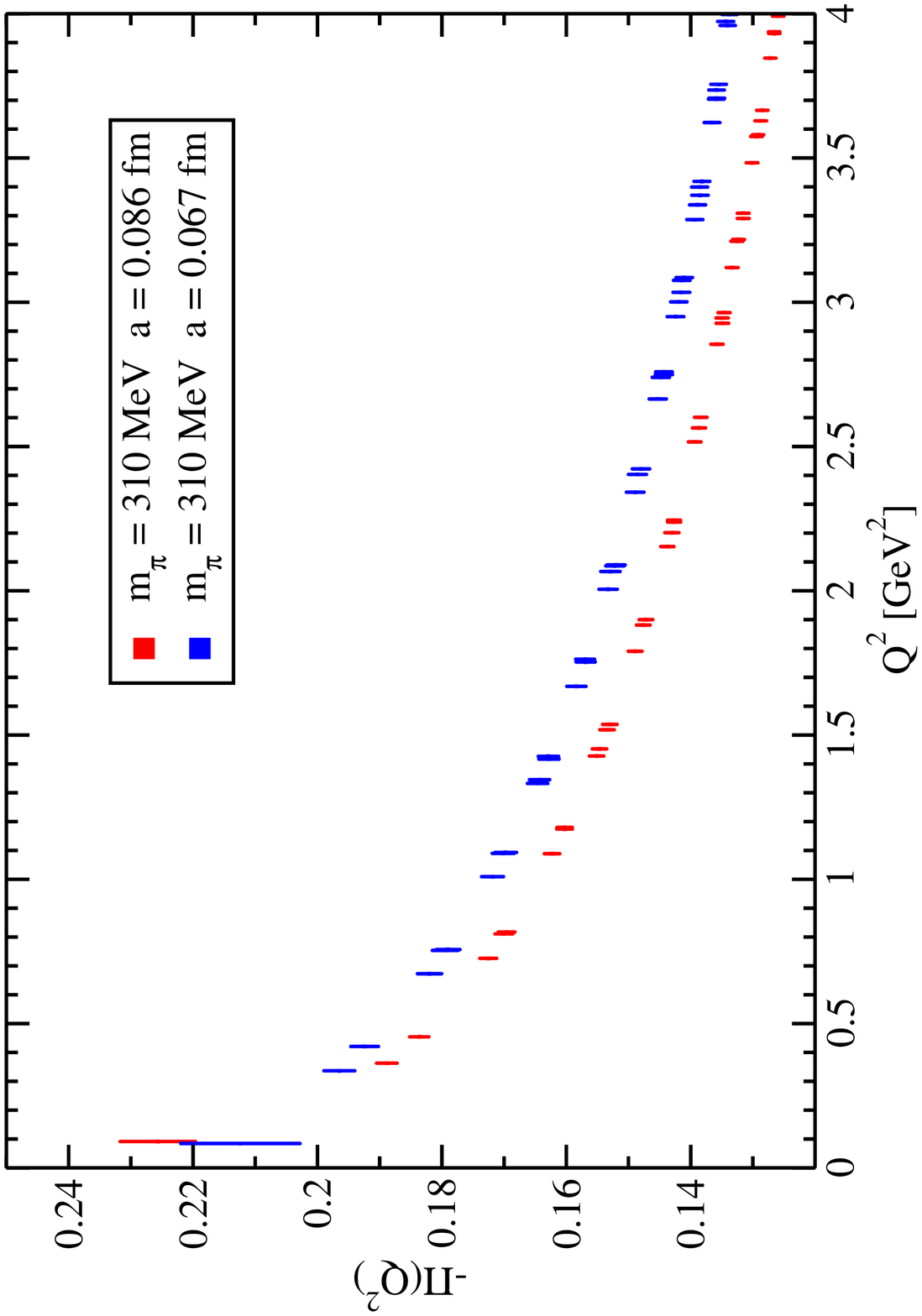}
\caption{Lattice spacing dependence of $-\pi(q^2)$.  The unrenormalized $-\pi(q^2)$ is shown.  The discrepancy
illustrates the ultra-violet subtraction required to renormalize $\pi$.}
\label{a_dep_1}
\end{minipage}
\hspace{4pt}
\begin{minipage}{210pt}
\includegraphics[width=\mywidth,angle=\myangle]{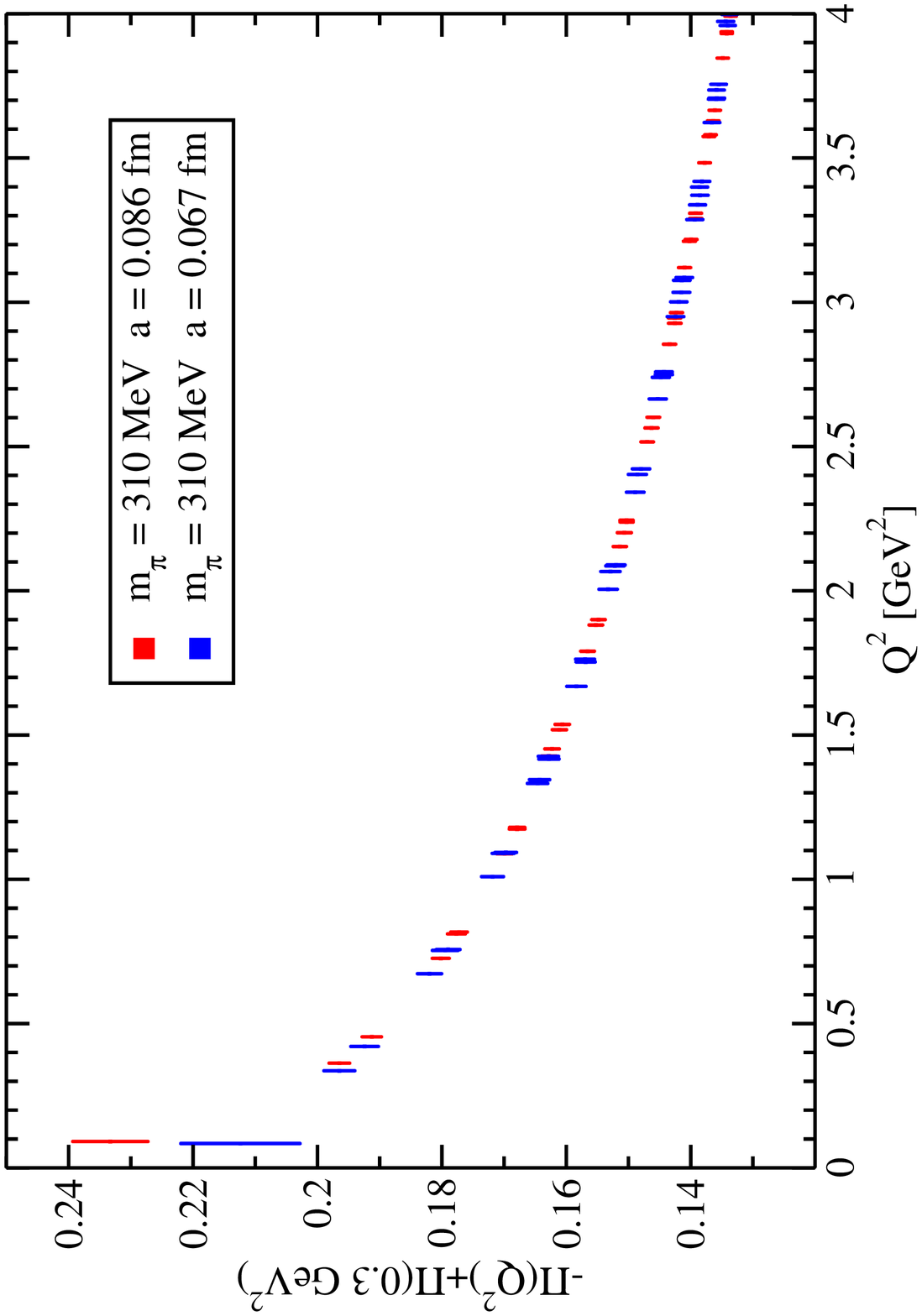}
\caption{Lattice spacing dependence of $-\pi(q^2)$ matched at $q^2=0.3~\mathrm{GeV^2}$.  With the exception of the lowest $q^2$ point, there is no noticeable
lattice spacing dependence.}
\label{a_dep_2}
\end{minipage}
\end{figure}
\begin{figure}
\begin{minipage}{210pt}
\includegraphics[width=\mywidth,angle=\myangle]{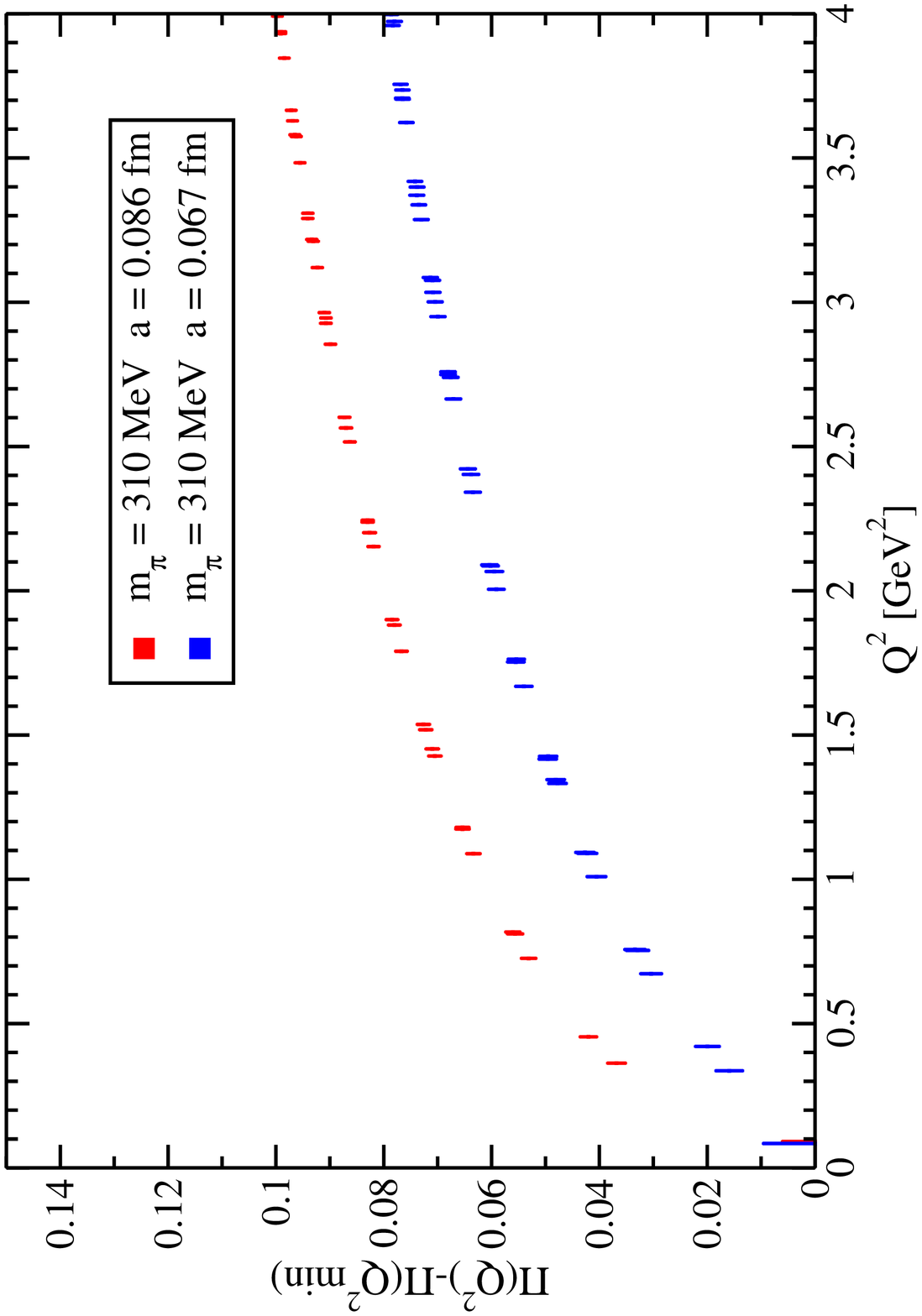}
\caption{Lattice spacing dependence of $\pi(q^2)-\pi(0)$.  The results have been renormalized at
the lowest value of $q^2$ at each lattice spacing to illustrate the effect of the subtraction.}
\label{a_dep_3}
\end{minipage}
\hspace{4pt}
\begin{minipage}{210pt}
\includegraphics[width=\mywidth,angle=\myangle]{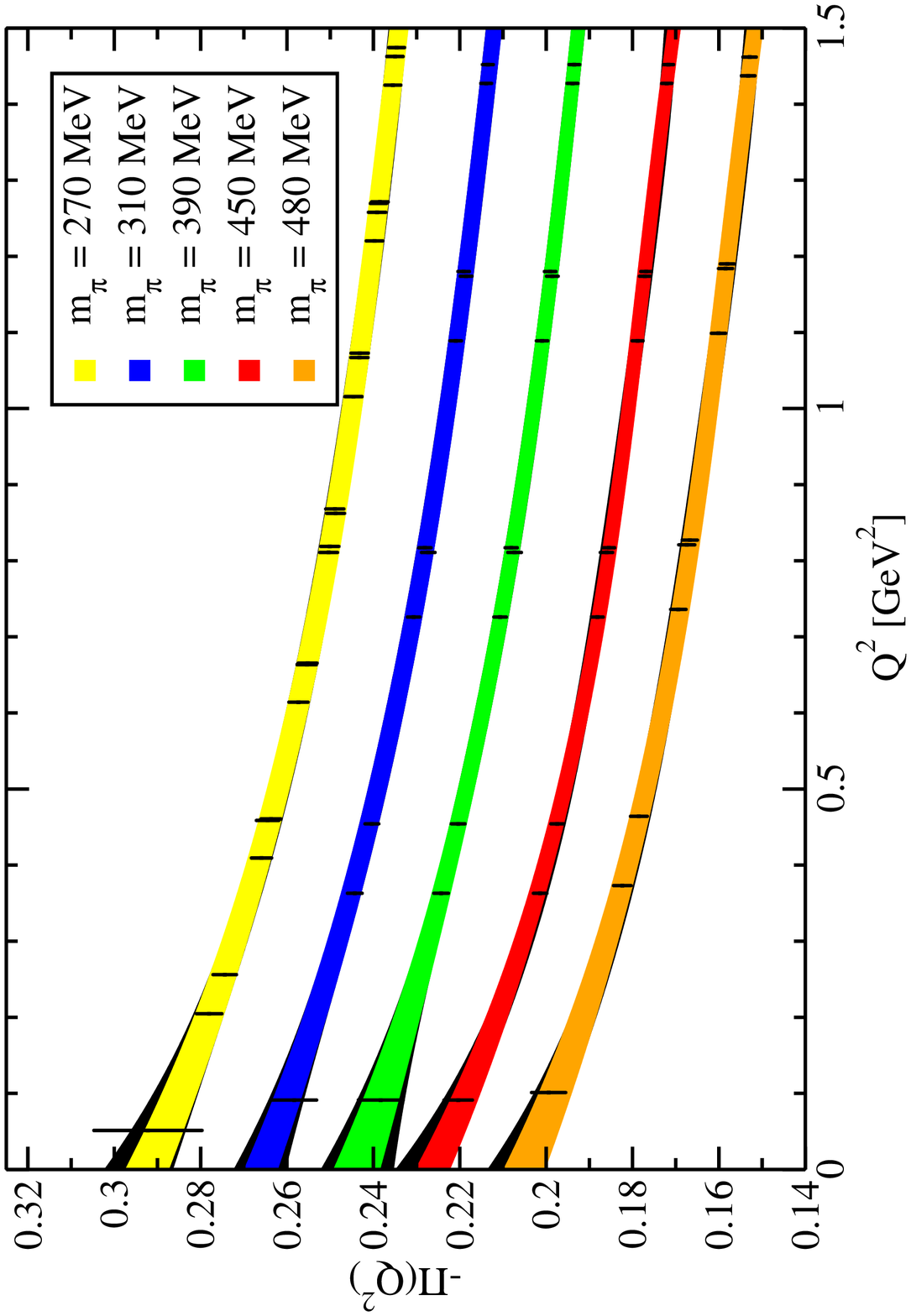}
\caption{Low $q^2$ extrapolation.  Each $\pi(q^2)$ has been fit to cubic (foreground) and quartic (background) functions of $q^2$, showing agreement for all but the smaller volume at $310~\mathrm{MeV}$ (not shown).}
\label{cubic}
\end{minipage}
\end{figure}

Now we examine the lattice artifacts in $\pi(q^2)$.  We have
calculated at two lattice spacings, $a=0.086~\mathrm{fm}$ and
$a=0.067~\mathrm{fm}$.  In both cases we have taken $m_\pi=310~\mathrm{MeV}$
and $L=2.1~\mathrm{fm}$.  Fig.~\ref{a_dep_1} shows the unrenormalized
results for $\pi(q^2)$ demonstrating the ultra-violet divergence
present without the subtraction.  In Fig.~\ref{a_dep_2} we perform the
subtraction, but at $q^2=0.3~\mathrm{GeV}^2$ rather than at $q^2=0$.
We see no noticeable lattice artifacts with the exception of the
lowest $q^2$ point.  Unfortunately the expression for $a_\mu$ in
Eq.~\ref{eq_amu} requires the subtraction at $q^2=0$.  This is shown
in Fig.~\ref{a_dep_3} where, as earlier, we subtract at the lowest $q^2$
point available in each calculation.  Again, the subtraction induces a
large difference between the results from the two lattice spacings, but we must remember that
$a_\mu$ is dominated by values of $q^2$ around $0.003~\mathrm{GeV}^2$.

To determine the extent to which the effects shown in
Figs.~\ref{v_dep_2} and \ref{a_dep_3} contribute to $a_\mu^{\mathrm{had}}$, we must
parametrize and fit $\pi(q^2)$ and extrapolate to $q^2=0$ in order to perform
the integral in Eq.~\ref{eq_amu}.  In Fig.~\ref{cubic} we show fits to
polynomials in $q^2$ with 4 terms (cubic) and 5 terms (quartic).  The
lattice results and corresponding curves are shifted vertically to
illustrate the quality of the fits and the nature of the extrapolation
to $q^2=0$.  Additionally, the calculation labeled
$m_\pi=310~\mathrm{MeV}$ refers to the larger $L=2.7~\mathrm{fm}$
calculation.  For all but one ensemble the cubic fit seems sufficient
to describe the lattice results.  The one exception is the
$m_\pi=310~\mathrm{MeV}$, smaller volume $L=2.1~\mathrm{fm}$
calculation (not shown), which requires the quartic term to
accommodate the observed low $q^2$ behavior.

\begin{figure}
\begin{minipage}{210pt}
\includegraphics[width=\mywidth,angle=\myangle]{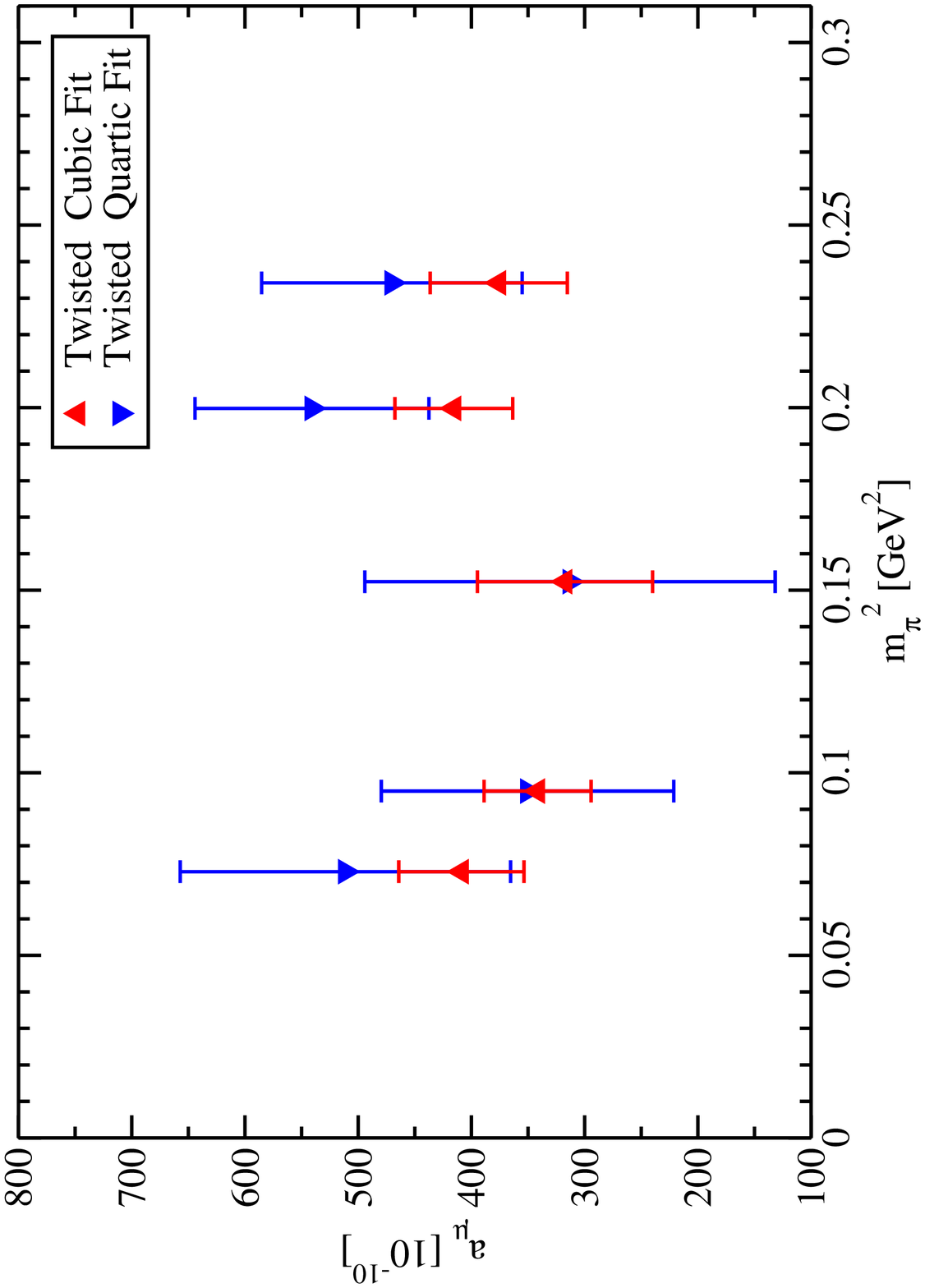}
\caption{Comparison of cubic and quartic fit results for $a_\mu^{\mathrm{had}}$.  There is general agreement between the
cubic and quartic fits for all calculations excluding the smaller volume at $m_\pi =310~\mathrm{MeV}$ (not shown).}
\label{sys_1}
\end{minipage}
\hspace{4pt}
\begin{minipage}{210pt}
\includegraphics[width=\mywidth,angle=\myangle]{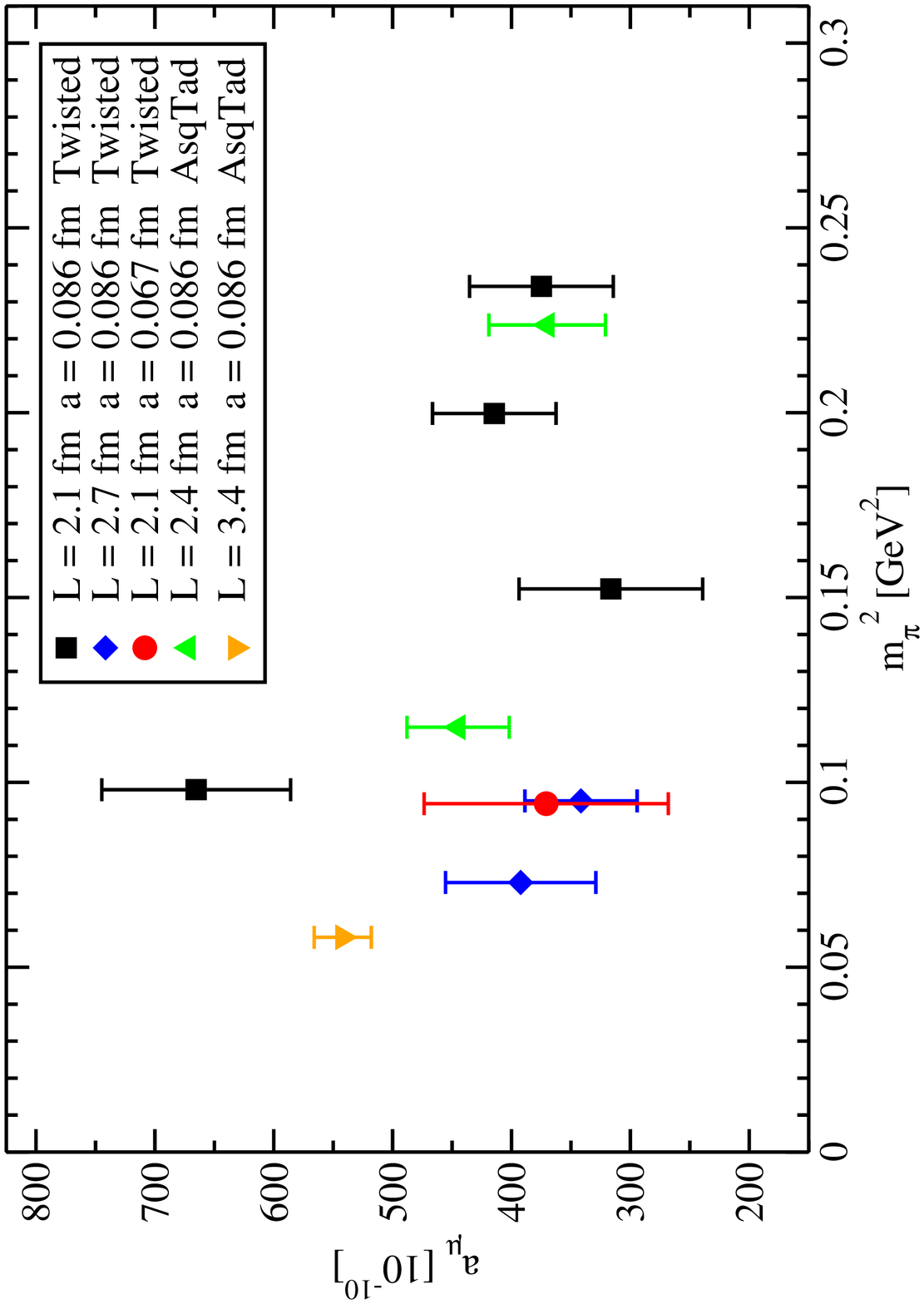}
\caption{Comparison of all full QCD calculations of $a_\mu^{\mathrm{had}}$.  The results of this work are shown along 
with the staggered results of \cite{Aubin:2006xv}.  Finite size effects and lattice artifacts are visible.}
\label{amu}
\end{minipage}
\end{figure}

Using the fits described above, we calculate $a_\mu^{\mathrm{had}}$ using
Eq.~\ref{eq_amu}.  Fig.~\ref{sys_1} shows the resulting values for the
five masses at $a=0.086~\mathrm{fm}$.  (Again the larger value of
$L=2.7~\mathrm{fm}$ is used for $m_\pi=310~\mathrm{MeV}$.)  We
note a clear consistency between the cubic and quartic fits as already
indicated in Fig.~\ref{cubic}.  Additionally, there is no discernible
quark mass dependence as implied by Figs.~\ref{mpi_dep_1} and
\ref{mpi_dep_2}.  In Fig.~\ref{amu}, we focus specifically on the
cubic fits and examine the $L$ and $a$ dependence of our results and we compare
to the only other full QCD calculation~\cite{Aubin:2006xv}\footnote{We will refer to the 
results in~\cite{Aubin:2006xv} as staggered and have taken the results
corresponding to the cubic fits in~\cite{Aubin:2006xv}.}.  First we
notice the quite large finite size effects and lattice artifacts as
anticipated in Figs.~\ref{v_dep_2} and \ref{a_dep_3}.  However, in
general we find consistency with the staggered results.  The agreement for
the largest staggered value of $m_\pi$ is quite clear.  The intermediate staggered
value is at a volume that is between our larger and smaller
volumes, and the result also lies between our two results.  This seems
quite consistent to within the systematic errors that we observe near
this value of $m_\pi$.  Finally $a_\mu^\mathrm{had}$ at the lightest staggered value of $m_\pi$ appears
to be a bit too high compared to our lightest value.  This might be
genuine quark mass dependence, but given the strength of finite size
effects and lattice artifacts that we observe, we find it difficult to
claim a strong quark mass dependence.  In fact the sign of the
discrepancy is consistent with both a finite size effect, which should
be universal, and also with our lattice artifacts, which, though not
universal, might still be indicative.

\section{Conclusions}

The current high precision determinations of the anomalous magnetic
moment, $a_\mu$, both from experiment and theory, indicate a small
discrepancy between Nature and the Standard Model.  The largest source
of error in the theory calculation of $a_\mu$ is the leading order
hadronic contribution.  We present a full QCD calculation of the
vacuum polarization and, in particular, of precisely this hadronic
contribution.  We perform calculations with dynamical maximally
twisted mass fermions with pion masses ranging from $480~\mathrm{MeV}$
to $270~\mathrm{MeV}$.  We observe both large finite size
effects and lattice artifacts but find general agreement with the only
other full QCD calculation.  This work presents the first full QCD
calculation of these effects and represents a first effort to begin to
calculate $a_\mu$ controlling for all sources of error.

\section{Acknowledgments}

We thank Karl Jansen and Carsten Urbach for their valuable suggestions and assistance
and Giancarlo Rossi and Gregorio Herdoiza for their comments on the proceedings.
We thank the John von Neumann Institute for Computing (NIC) for computing
resources and the staff at the J{\"u}lich Supercomputing Center and DESY Zeuthen
Computing Center for their support.  This work has been supported in part by the DFG
Sonder\-for\-schungs\-be\-reich/Transregio SFB/TR9-03 and the DFG project Mu 757/13.

\bibliography{gm2.bib}

\end{document}